\documentclass[modern]{aastex61}
\usepackage{times}
\usepackage{geometry}
\usepackage[utf8]{inputenc}
\usepackage{enumitem,amssymb}
\usepackage{ragged2e}
\newlist{thematic}{itemize}{8}
\setlist[thematic]{label=$\square$}

\usepackage{floatrow}

\usepackage{tcolorbox}
\definecolor{azure}{RGB}{240, 255, 255}

\begin{document}
\raggedright
\begin{center}
\huge
Astro2020 Science White Paper \linebreak
\Large
\vspace{0.2cm}
The trail of water and the delivery of volatiles to habitable planets \linebreak
\end{center}
\normalsize

\noindent \textbf{Thematic Areas:}  Planetary Systems, Star and Planet Formation 

\vspace{0.2cm}
\textbf{Principal Author: }

Name: Klaus M. Pontoppidan
 \linebreak						
Institution:  Space Telescope Science Institute
 \linebreak
Email: pontoppi@stsci.edu
 \linebreak
Phone:  (410) 338-4744
 \linebreak
 
\textbf{Co-authors:} 
  \linebreak
Andrea Banzatti (University of Arizona), Edwin Bergin (University of Michigan), Geoffrey A. Blake (Caltech), Sean Brittain (Clemson University), Maryvonne Gerin (Observatory of Paris), Paul Goldsmith (JPL), Quentin Kral (Observatory of Paris), David Leisawitz (NASA GSFC), Dariusz Lis (Caltech), Melissa McClure (University of Amsterdam), Stefanie Milam (NASA GSFC), Gary Melnick (CfA), Joan Najita (NOAO), Karin {\"O}berg (Harvard), Matt Richter (UC Davis), Colette Salyk (Vassar College), Martina Wiedner (Observatory of Paris), and Ke Zhang (University of Michigan).

\vspace{1cm}

\begin{tcolorbox}[width=\textwidth,colback={azure},colbacktitle=yellow,coltitle=blue]
\begin{center}
\textbf{Abstract}\\
\end{center}
Water is fundamental to our understanding of the evolution of planetary systems and the delivery of volatiles to the surfaces of potentially habitable planets. Yet, we currently have essentially no facilities capable of observing this key species comprehensively. With this white paper, we argue that we need a relatively large, cold space-based observatory equipped with a high-resolution spectrometer, in the mid- through far-infrared wavelength range (25-600~$\mu$m) in order to answer basic questions about planet formation, such as where the Earth got its water, how giant planets and planetesimals grow, and whether water is generally available to planets forming in the habitable zone of their host stars.
\end{tcolorbox}

\section*{Motivation} 

\noindent {\bf Water is critical for the emergence and evolution of life as we know it.} Life on Earth is dependent on water, with the addition of carbon, nitrogen, and other minor elements \citep{Chopra10}. During the process of planet formation, the abundance and phase (solid or gas) of water traces the flow of volatile elements toward their ultimate incorporation into potential biospheres \citep{Marty12}. Understanding how the ingredients for life are delivered to exoplanets requires new observational constraints on the distribution of water, in particular during the process of planet formation. As our understanding of the flow of volatiles to protoplanetary disks, and to exoplanets, grows tighter, interest in water will grow. By the late 2020s, there will likely be great interest in understanding this {\it trail of the water}.

\vspace{0.2cm}
\noindent {\bf Water is thought to play a critical role in the growth of planets.} The recent Atacama Large Millimeter Array (ALMA) images of intricate dust structures \citep{Andrews18} provide dramatic evidence of the importance of the physics of solids in protoplanetary disks. As solid cores are needed to form giant planets, the presence of a massive ice reservoir may profoundly affect the architecture of exoplanetary systems \citep{Ida04,Raymond04} and the prevalence of water worlds \citep{Zeng18}. Indeed, the dust seen by ALMA is likely icy, with a mass and volume dominated by water ice. If so, planetesimal formation may only occur beyond the water snowline \citep{Drazkowska17}. Thus, understanding the observed structure of solids and the emergence of both rocky and giant planets requires observations of the water content in dust and gas in protoplanetary disks. 

\vspace{0.2cm}
\noindent {\bf Water is intimately linked to the composition of exoplanet atmospheres.} Giant planets, planetesimals, as well as comets and Kuiper belt objects form in a rich chemical environment where primordial water and other volatiles mix with new complex chemistry to create a broad diversity of planetary systems. For instance, one of the central links between planet composition and disk composition that is currently driving efforts in both planet formation and exoplanet communities is the elemental C/O ratio. Since water is the dominant carrier of oxygen, it is a driver of the C/O ratio and the relative partition of carbon and oxygen between the gas and solid phases. Indeed, giant planets have C/O ratios set by their birth location and they may, depending on core-envelope mixing and planetesimal accretion, carry this information to later stages \citep{Oberg11}. There is presently an extensive industry attempting to determine C/O ratios in planetary atmospheres \citep[e.g.,][]{Madhusudhan12,Konopacky13,Line14,Dawson18}, but also in protoplanetary disks \citep[e.g.,][]{Kama16,Bergin16,Cleeves18}. It is also thought that many chemical signatures seen in primordial Solar System materials originate in the gas-rich protoplanetary disk phase \citep{Busemann06,Mumma11,Simon11}. 

\pagebreak
\section*{How water is observed}

\begin{figure}[ht]
\begin{center}
\includegraphics[width=13cm]{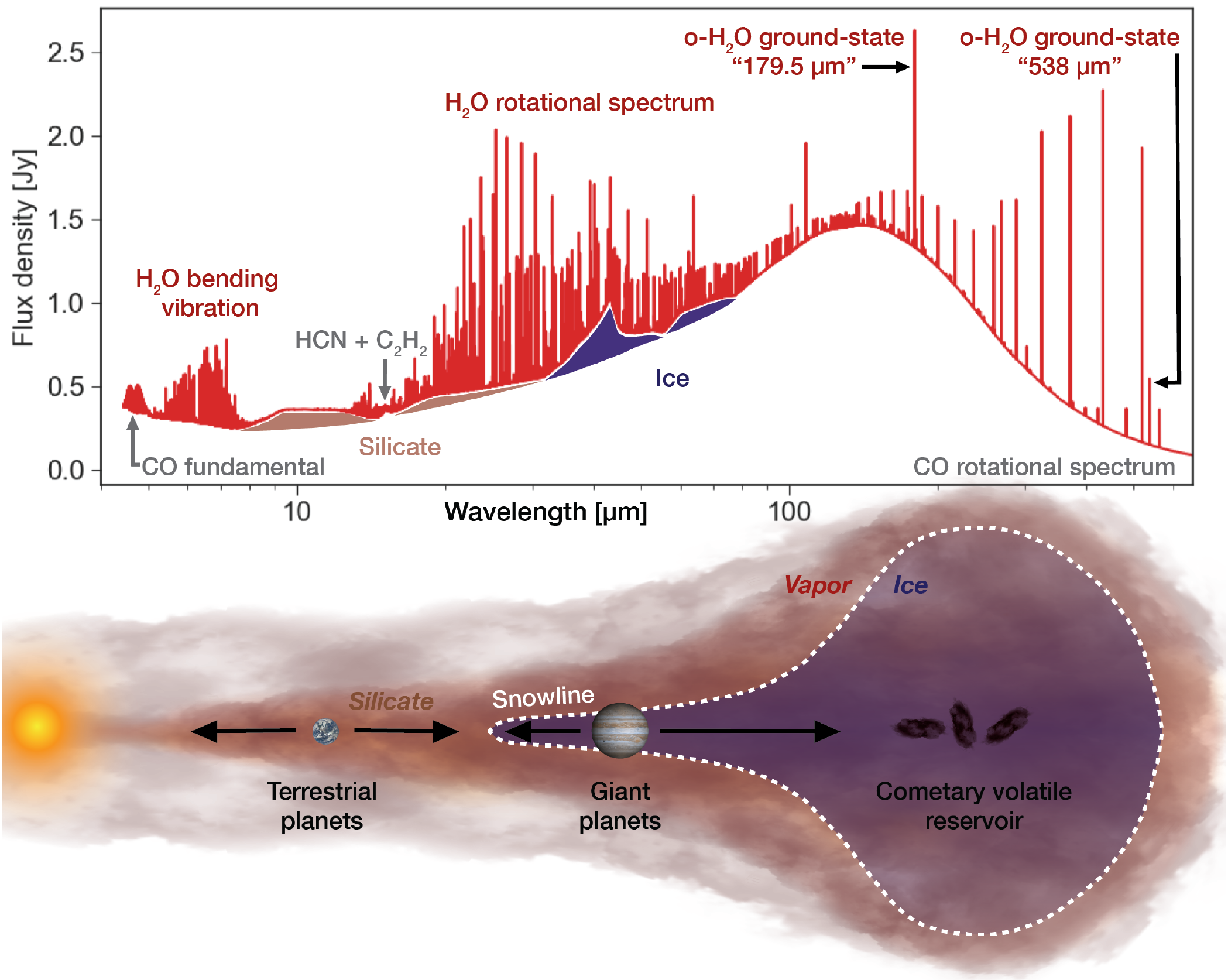}
\end{center}
\caption{\it The broad infrared wavelength range in protoplanetary disks is in many cases dominated by strong emission from water lines when observed at high spectral resolving power. This plot shows a model of a typical disk around a solar-mass star at a distance of 125\,pc, with line strengths fitted to {\it Spitzer} and {\it Herschel} spectra using a two-dimensional radiative transfer model \citep{Blevins16}. The model is rendered at a resolving power of $R=50\,000$, and viewed at an inclination of 45 degrees. Also visible are the strong far-infrared emission bands of crystalline water ice.}
\label{disk_model}
\end{figure}

To understand the role of water, it is necessary to measure the mass and spatial distribution of water relative to hydrogen in all of the water reservoirs in disks. Figure \ref{disk_model} shows the spectral tracers of water in a typical protoplanetary disk.

{\bf Water vapor} can generally be observed via a large number of rotational and ro-vibrational transitions throughout the infrared to submillimeter range (2-1000\,$\mu$m). We know water is abundant in disks, both from theoretical expectations, as well as from Spitzer and Herschel detections \citep[e.g.,][]{Pontoppidan10a,Carr11,Riviere-Marichalar12}. Transitions at longer wavelengths have a tendency to trace lower excitation temperatures. Because the Earth's atmosphere absorbs strongly in water transition with upper level energies below $\lesssim$700\,K, cold water is not accessible from the ground, or even from airplanes. Conversely, water transitions excited in hot gas ($T\gtrsim 700\,$K) can be observed from the ground, though with some difficulty \citep[e.g.,][]{Carr04, Pontoppidan10b, Salyk15, Banzatti17, Adams18}. The sensitivity to water transitions from the ground is generally low, given the high backgrounds from warm telescopes. 
\vspace{0.2cm}

There is some demonstrated potential for detecting the 321~GHz warm water line from the ground ($10_{2\,9}\rightarrow 9_{3\,6}$, $E_{\rm upper}=1861\,$K), but probably only for a few of the brightest targets with current ALMA sensitivities \citep{Carr18}. Consequently, observing the vast majority of the water vapor mass in protoplanetary disks requires a sensitive space observatory. The James Webb Space Telescope (JWST) will observe hot water in many disks, and thereby make great strides in our understanding of water within $\sim 1$~au. Access to wavelengths longer than the 28\,$\mu$m limit of JWST is needed to observe any water lines with  $E_{\rm upper}\lesssim 800\,$K. 

\begin{figure}[ht]
\begin{center}
\includegraphics[width=4.3cm,angle=270]{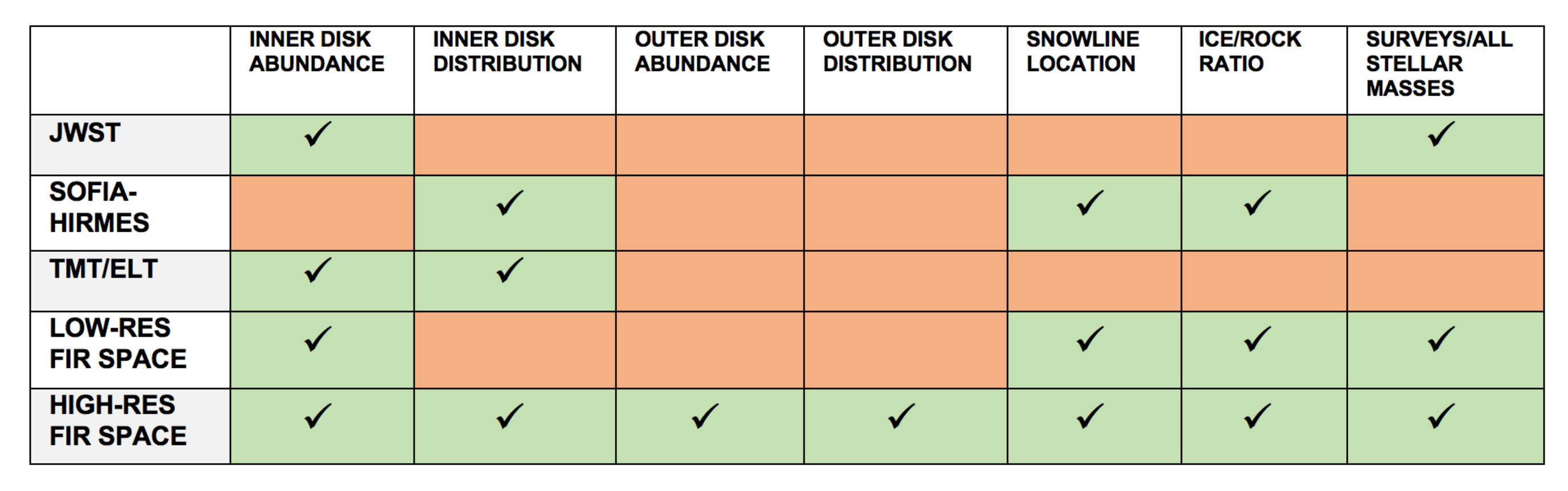}
\end{center}
\caption{\it Tick marks indicate a potential of existing and planned observatories for observing reservoirs of water in protoplanetary disks. Filling this table will address 1) how the water snowline location evolves during planetesimal and planet formation, 2) what are the water abundances inside and outside the snowline at different evolutionary stages, and 3) how does the ice/rock ratio evolve as a function of radius. While ground-based and mid-infrared facilities can observe water in limited forms, only a large far-infrared space telescope is capable of comprehensively observing water in all its forms.} 
\label{obs_summary}
\end{figure}

Typical integrated water line fluxes across the energy range for a disk around a solar-mass young star in the nearest star-forming regions ($\sim$150\,pc) are $10^{-18}-10^{-17}\,\rm W\,m^{-2}$. The ground-state line at 538\,$\mu$m tends to be weaker at $10^{-19}\,\rm W\,m^{-2}$. Investigating water in disks around low-mass stars, or at larger distances (e.g., Orion), requires line sensitivities in excess of $\sim 10^{-19}\,\rm W\,m^{-2}$.

\vspace{0.2cm}
{\bf Water ice} can be observed at mid-infrared wavelengths through solid state vibrational transitions (primarily at 3, 6 and 11\,$\mu$m), or via phonon modes at $\sim$43-47 and $\sim$62\,$\mu$m. The mid-infrared ice bands can only be observed in absorption or scatting in disks under special circumstances \citep{Pontoppidan05,Honda09} because heating dust grains to the temperatures where the bands would appear in emission will also desorb any ice. The far-infrared features are {\it unique} as they can be excited at relatively low temperatures, and therefore be seen in emission from the bulk water ice reservoir \citep{McClure15}. The strongest 43\,$\mu$m ice feature was not observable within the bandpasses of either {\it Spitzer} or {\it Herschel}, but would be readily accessible to a space telescope of modest size with sensitivities of a few mJy ($1\sigma$) in the 35--100\,$\mu$m range \citep{Kamp18}. Figure \ref{obs_summary} summarizes how different observational facilities can observe the various properties of water in protoplanetary disks.

\section*{What cannot presently be done}

We presently cannot observe, or unambigously characterize, most of the water in planet forming regions. Since water represents the bulk volatile reservoir, this deficit will restrict our understanding of the delivery of volatiles to planets. ALMA is showing evidence for a rich and active chemistry \citep[e.g.,][]{Bergin18} in the carbon and nitrogen elemental pools. This work is ground-breaking but cannot directly constrain the O in the C/O ratio for either the solids or the gas, as a direct, and robust, tracer of cold water and ice is lacking. While ALMA can probably detect warm water in a very small number of cases (e.g. a tentative detection by \cite{Carr18}), it cannot provide broad access to the cold water reservoir. JWST-MIRI can access water vapor emission from the hot disk surface inside the snowline, but cannot access the dominant cold water reservoir ($T < 200$\,K). Furthermore, JWST cannot spectrally (or spatially) resolve the hot water lines to constrain the location of the snowline, as the relevant lines have widths in the range 10-50\,$\rm km\,s^{-1}$ -- significantly less than the best JWST resolving powers of 75-200\,$\rm km\,s^{-1}$. Ground-based mid-infrared spectroscopy could spectrally resolve lines from hot water, but is presently only sensitive to the brightest systems. A high-resolution, mid-infrared spectrometer on a 30m class telescope could make important contributions to our understanding of the hot water inside the snowline. To measure the total water content in a representative sample of planet-forming disks, we will need full access to the mid- to far-infrared spectrum of water from a space-based platform.

\begin{figure}[h]
\centering
\floatbox[{\capbeside\thisfloatsetup{capbesideposition={right,top},capbesidewidth=3.7cm}}]{figure}[\FBwidth]
{\caption{\it \small Overview of the sensitivity to the water mass at all phases and evolutionary stages compared to the sensitivities of various previous and planned observatories (SPICA, JWST, Herschel, and Origins). For each case we provide the maximum range of expected values based on models and/or observations shown in purple rectangles.}\label{fig:test}}
{\includegraphics[width=11cm]{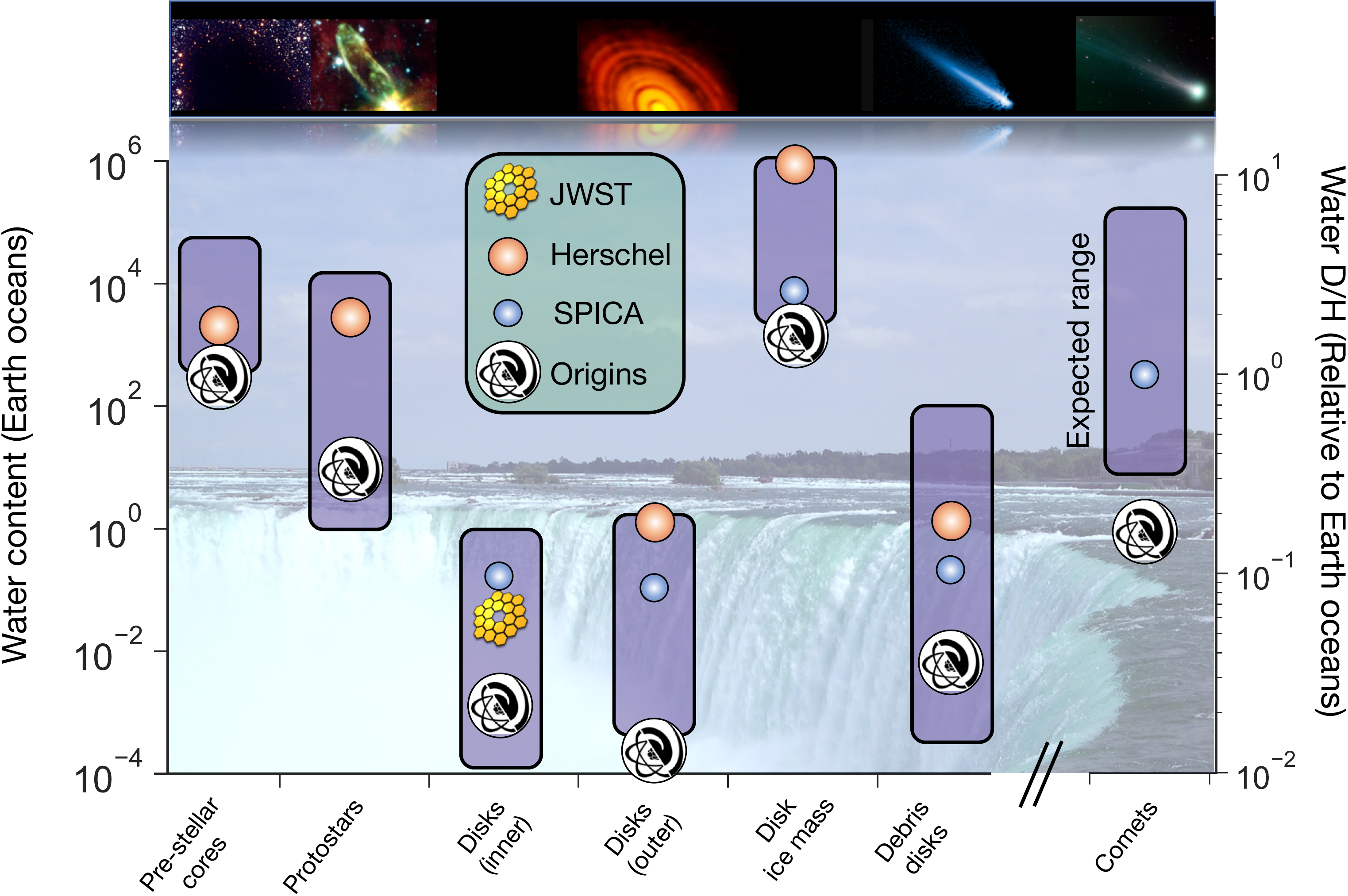}}
\label{sensitivity_figure}
\end{figure}

\section*{Recommendations}

In order to accomplish the goal of tracing bulk water vapor and ice at all temperatures, from 10-1000~K, during planet formation, it is necessary to have access to {\bf sensitive line spectroscopy at high spectral resolution from mid-infrared to the far-infrared wavelengths (3-600\,$\mu$m)}. There is a sliding scale of capabilities, where some aspects of planet-forming water can be explored by less capable facilities (lower sensitivity, limited bandpass, and lower spectral resolution). Important and focused contributions will be made by observatories currently under development, including JWST, SOFIA-HIRMES, and ELT/TMT. JWST will measure the relative amount of hot water in the inner disk surface inside the snow line, for a significant number of disks, while SOFIA-HIRMES will measure the distribution of water out to, and including, the snow line for a smaller number of disks due to sensitivity limitations. These will be important contributions in the next decade (see Figure \ref{sensitivity_figure}). 

\vspace{0.2cm}
{\bf Recommendation 1 -- water is a key tracer of planet formation processes:} Observing water vapor at all temperatures, as well as water ice, in protoplanetary disks will be crucial for understanding the development of habitable planets and the formation of giant planets in the next decade. As the composition of exoplanetary atmospheres becomes better established, it will be crucial to link them to observations of volatiles during the process of planet formation. 

\vspace{0.2cm}
{\bf Recommendation 2 -- sensitive mid- to far-infrared spectroscopy is needed to detect water at all temperatures and phases:} While hot water ($T\gtrsim 500\,$K) can be traced at 5--28\,$\mu$m, observing cooler water, and in particular water tracing the snow line, requires sensitive spectroscopic access to the far-infrared (25-600\,$\mu$m). Indeed, the far-infrared is the most effective wavelength region as it has the potential to trace both hot and cold water vapor, as well as bulk ice. Efficiently measuring water vapor at all temperatures in nearby (150\,pc) protoplanetary disks around solar-mass stars requires line sensitivities of at least 10$^{-19}$\,W\,m$^2$ (10$\sigma)$. From space, the minimum required wavelength coverage includes at least 25-200\,$\mu$m, which includes strong water transitions with upper level energies between 2000 and 114\,K. The addition of the lowest-lying ground-state line of ortho-water at 538\,$\mu$m, with a relative upper level energy of 27\,K, provides access to the coldest water at 10-20\,K. 

\vspace{0.2cm}
{\bf Recommendation 3 -- high spectral resolving power is needed to localize water vapor:} High spectral resolving powers of at least $25\,000-50\,000$ are needed to maximize the line-to-continuum ratio and therefore the contrast to the bright dust far-infrared continuum present in protoplanetary disks. These resolving powers will also measure the profiles of almost all water lines tracing gas out to, and including, the snow line. Even higher spectral resolving powers in excess of $R\sim 200\,000$ are required to precisely locate the coldest gas in outer disks using kinematic tomography of the ground-state water lines at 179.5 and 538\,$\mu$m (as well as other lines of small molecules in protoplanetary disks, such as NH$_3$ and other hydrides). Observations of the far-infrared water ice band at 43~$\mu$m requires considerably lower resolving powers of a few hundred, but need spectro-photometric calibration at the 1\% level to detect ice species other than water \citep{Kamp18}.

\vspace{0.2cm}
Ultimately, a large, cold space telescope with access to the 25--600\,$\mu$m range equipped with a high-resolution spectrometer ($R=25\,000-200\,000$) is required to explore all water reservoirs in planet-forming systems around stars of all masses. The {\it Origins Space Telescope} concept is designed to meet the requirements for comprehensively observing water at all temperatures during the process of planet formation.

\pagebreak

\bibliographystyle{aasjournal}
\bibliography{ms}

\end{document}